\documentstyle[11pt,paspconf,epsf]{article}

\def\etal{{\it et al.\ }}
\def\eg{{\it e.g.\ }}

\begin{document}

\title{Bars within Bars in Galaxies}
\author{Witold Maciejewski\altaffilmark{1} and Linda S. Sparke}
\affil{Department of Astronomy, University of  Wisconsin, \\
475 N. Charter St., Madison, WI 53706-1582}
\altaffiltext{1}{also Max-Planck-Institut f\"ur Astronomie, K\"onigstuhl 17,
D-69117 Heidelberg, Germany, {\tt witold@mpia-hd.mpg.de}}

\begin{abstract}
Inner regions of barred
disk galaxies often include asymmetrical, small-scale 
central features, some of which are best described as secondary 
bars. Because orbital timescales in the galaxy center are short, 
secondary bars are likely to be dynamically decoupled from the 
main kiloparsec-scale bars. We found that non-chaotic 
multiply-periodic 
particle orbits can exist in 
potentials with two dynamically decoupled bars.
Stars trapped around 
these orbits could form the building blocks for a long-lived, 
doubly-barred galaxy. 
A self-consistent secondary bar appears to induce formation
of inner gaseous rings rather than shocks in gas flow.
\end{abstract}

\section{Basic concepts}
A number of nearby  barred
galaxies shows isophotal twists within the central few hundred parsecs: 
the signature of a secondary bar (Erwin \& Sparke 1998, this volume). 
The inner bars appear to be oriented randomly with respect to the larger
bars (see \eg Friedli 1996),
as expected if they are dynamically
distinct subsystems. Their presence in infrared images suggests they may
contain old stars, and do not consist purely of young stars and gas.

Understanding the dynamical state of these inner bars --- whether
they are long-lived or transient, and what orbits the stars and gas follow
within them --- requires knowledge not only of the size, shape and
strength of the bar, but also of whether its orientation is fixed with
respect to the main bar, or rotates about the galaxy center with some
nonzero pattern speed. Within about 100 pc of a galactic 
center, orbital times are at least an order of magnitude
less than those at a few kiloparsecs, thus a dynamically 
decoupled inner bar is likely to rotate faster than the
outer structure. The entire pattern is then not steady in any
reference frame. Orbits in the doubly barred potential do not have a
conserved integral of motion, and in principle they might all be
chaotic, exploring large regions of phase space. If the orbits
are mostly chaotic, it is unlikely that such a system could be
{\it self-consistent},  
so that the average density of all the stars on their
orbits in the time-varying potential adds up exactly 
to give rise to the potential in which they move.
Nevertheless, secondary nuclear bars have been seen to form in numerical
simulations involving gravitating particles together with dissipative `gas
clouds' (see \eg Friedli 1996).

How can potentials
including two independently rotating bars maintain themselves as
gravitating systems? What are the conditions under which a gravitationally 
self-consistent double-bar structure
could exist? We approach these questions by 
considering models that include two rigid bars, which rotate
at two constant, incommensurable pattern speeds, with other galaxy model
components defined after Athanassoula (1992).
Among these, we look for 
models that are close to being self-consistent,
i.e. those supporting orbital families capable of 
hosting sets of particles that together recreate the
assumed time-dependent density distribution. In the stationary 
potential of a single bar, particles on a closed periodic orbit 
move along it, always staying on the same curve, which is the orbit
--- the backbone of a steady potential.
Orbits in a double-barred galaxy
will generally not be closed in any uniformly rotating frame of reference,
since particles there undergo two forcing actions with 
non-commensurable frequencies. 

We want to extend the definition of an orbit, in order to find 
closed curves which can serve as backbones of a non-steady, 
doubly barred system. We search for curves that return to their
original positions every time the bars come back to the same relative 
orientation. We call these curves {\it the loops}:
a particle that begins its orbit from a position along a given
loop returns to another point on the same loop
after the bars have realigned. Loops
change their shape as the bars rotate through each other.
Particles trapped around loops that stay aligned with the bars in their
motion could form the building blocks for a long-lived, self-consistent,
doubly-barred galaxy. 
In general, every time the bars 
realign, the particle takes a different position on the loop; eventually 
it fills the whole loop. Thus we only need to know the 
initial conditions for one particle to recover the whole loop. The initial 
guess can be taken from the epicyclic approximation, in which the
loop is a set of particles with the same guiding radius
(Maciejewski \& Sparke 1997).

\section{Loops supporting a doubly barred galaxy}
\begin{figure}
\plotfiddle{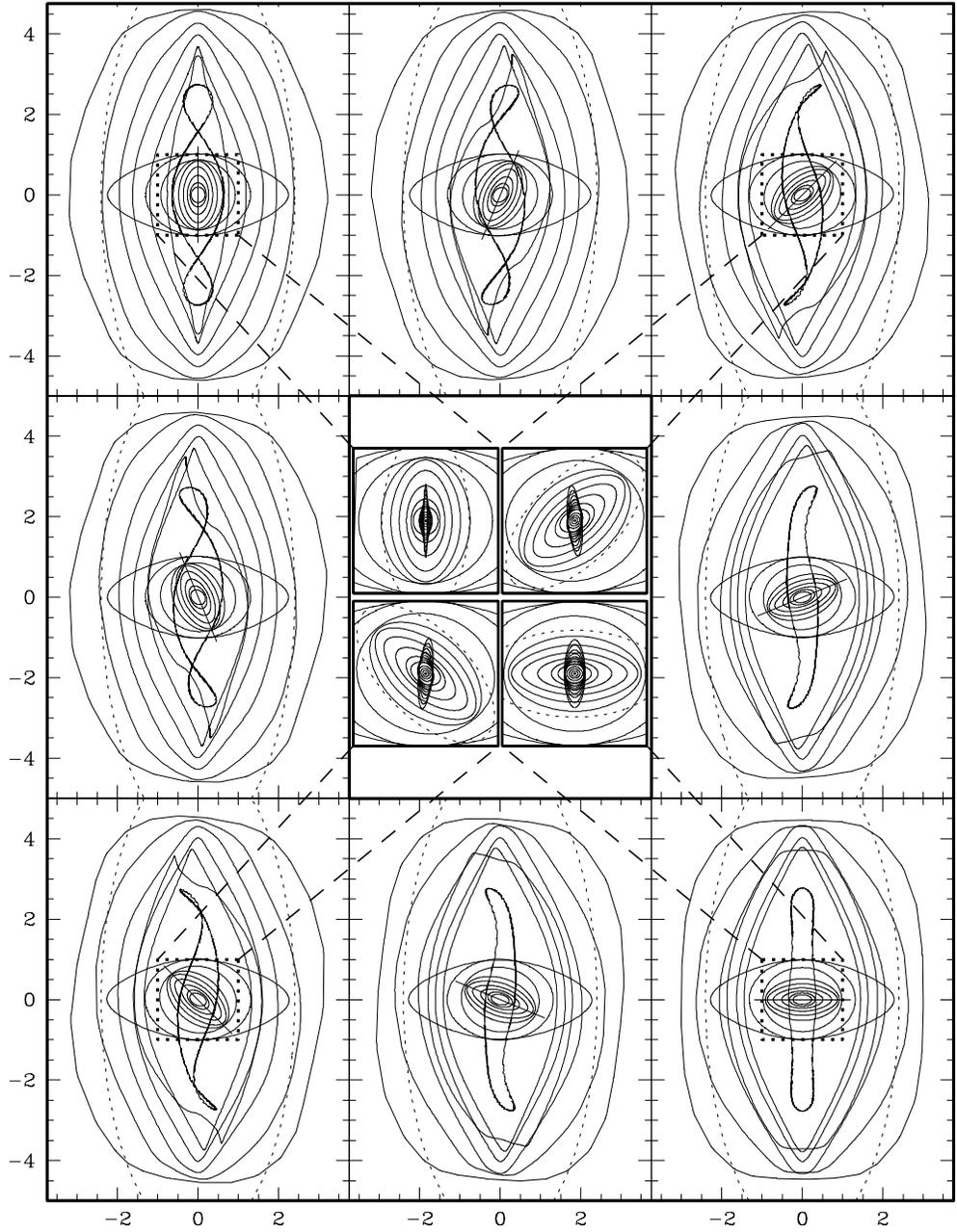}{10cm}{0}{90}{90}{-290}{-330}
\vspace{6cm}
\caption{Loops in a doubly barred galaxy, at different relative 
positions of the bars, displayed in the reference
frame of the main bar, which remains vertical. The outer bar
is outlined in dashed and the major axis of the secondary bar is
marked by a straight line. The sequence follows along outer panels
clockwise, with central regions magnified on inner panels, where the 
secondary bar is outlined in dashed. Units on the axes are in kiloparsec.}
\end{figure}
In our models, the most important loop families occupy the same
parts of phase space as the $x_1$ and $x_2$ orbits in the single
bar. The loop family corresponding to 
the $x_1$ orbital family in the main bar (the $x_1$ loops) often
breaks into an inner part $x_{1i}$ and an outer one $x_{1o}$;
there appear to be no stable loops at intermediate radii. In Figure 1,
loops belonging to the $x_{1o}$ family are elongated vertically,
and located outside a loop that intersects itself at bars aligned (the 
upper-left panel). The $x_{1i}$ loops, for clarity displayed on the
inner panels only, reside inside the secondary bar and also
remain vertical. The loop family that at large radii 
corresponds to the $x_2$ orbital family of the main bar
changes continuously with decreasing radius;
it begins to follow the secondary bar in its motion,
so it eventually corresponds to its $x_1$ orbital family.
We will call this family the $x_2$ loop family: like the $x_2$ 
orbits in a single bar, these loops do not extend all the way to the 
center. This family is represented by a set of horizontal loops on the
lower-right panel of Figure 1, and by corresponding sets on other panels.
No loop family corresponding to the $x_2$ orbital family in the
secondary bar has been found. 

The loops supporting the secondary bar are those $x_2$ loops which rotate 
smoothly with its figure. They change axial 
ratio as the bars rotate, and lead or trail the figure of the
secondary bar: a self-consistent secondary bar is likely to pulsate 
and accelerate as it revolves inside the
main bar. We find that the size of 
a self-consistent secondary bar is approximately
limited by the maximum extent of the $x_2$ orbits along the main bar's 
major axis. Since a strong secondary bar can easily disrupt orbits 
supporting the main bar, 
its mass is limited by the requirement that a substantial
part of the main bar is supported by the $x_1$ loops, so that the galaxy
remains doubly barred. Even then, the $x_{1o}$ loops are strongly 
influenced by the motion of the secondary 
bar (they get rounder when the bars are orthogonal), and
we were not able to find any $x_1$ loops supporting
the inner part of the main bar when the secondary bar is perpendicular
to it. A partial support to the inner region of the large bar is given by
another loop family (we call it the $b_T$ family), represented
in Figure 1 by the 
curve which intersects itself at bars aligned (upper-left panel),
and remains aligned with the main bar.

\section{Gas flows in doubly barred galaxies}
We model the flow of isothermal gas
using the CMHOG PPM code (written by James M. Stone, modified
by Piner \etal 1995)
on a staggered grid in planar polar coordinates 
($\Delta R \simeq R\Delta\varphi$), which gives excellent
resolution of circumnuclear phenomena near the grid center.
Initially, uniformly distributed gas is in circular motion;
the potential of the primary bar is then smoothly imposed.
The secondary bar is introduced after the flow in a single bar has 
been stabilized.

The interior of the nuclear ring, formed by gas flow in a single bar 
(see Piner \etal 1995), assumes an elliptical shape in the presence of a
secondary bar. The interior begins to rotate with the secondary bar, and 
eventually an elliptical ring forms around it.
A circular disk develops inside the ring; this disk turns
into another, circular ring by the end of the simulation. 
Neither of the rings
forms a shock --- the gas moves along the rings, and inside the inner 
circular ring it is in almost perfect circular motion. 
The lack of shocks within the secondary bar (unlike in the primary bar, 
where they manifest themselves as dust lanes)
can be explained by recalling, that 
because of the limitations imposed by the orbital structure,
this bar extends only about half-way to its corotation. Athanassoula 
(1992) found that in this case the 
dust lanes curl around the bar and start forming a ring. Therefore
some self-consistent doubly barred galaxies may lack secondary
shocks or dust lanes, and may not induce strong gas inflow to the
galaxy center. A better exploration of possible potentials
is needed in order to reach firmer conclusions.

\end{document}